\documentclass[letterpaper,twocolumn,prl,aps,superscriptaddress,amsmath,amssymb,floatfix]{revtex4-1}
\DeclareMathAlphabet{\mathcal}{OMS}{cmsy}{m}{n}

\usepackage[latin9]{inputenc}
\usepackage{color}
\usepackage{verbatim}
\usepackage{float}
\usepackage{amsmath}
\usepackage{amssymb}
\usepackage{physics}
\usepackage{graphicx}
\usepackage{esint}
\usepackage[unicode=true,
 bookmarks=true,bookmarksnumbered=false,bookmarksopen=false,
 breaklinks=false,pdfborder={0 0 1},backref=false,colorlinks=true]
 {hyperref}
\hypersetup{
 linkcolor=magenta,urlcolor=blue,citecolor=blue,pdfstartview={FitH},hyperfootnotes=false}
\makeatletter

\usepackage{epsfig}\usepackage{amsfonts}

\usepackage{soul}

\newcommand{\Zcite}[1]{\textcolor{red}{[cite]}}

\makeatother

\begin{document}

\title{Optimized detector tomography for photon-number resolving detectors with hundreds of pixels}

\author{Dong-Sheng Liu}
\affiliation{CAS Key Laboratory of Quantum Information, University of Science and Technology of China, Hefei 230026, China}
\affiliation{Center For Excellence in Quantum Information and Quantum Physics, University of Science and Technology of China, Hefei 230026, China}

\author{Jia-Qi Wang}
\affiliation{CAS Key Laboratory of Quantum Information, University of Science and Technology of China, Hefei 230026, China}
\affiliation{Center For Excellence in Quantum Information and Quantum Physics, University of Science and Technology of China, Hefei 230026, China}

\author{Chang-Ling Zou}
\affiliation{CAS Key Laboratory of Quantum Information, University of Science and Technology of China, Hefei 230026, China}
\affiliation{Center For Excellence in Quantum Information and Quantum Physics, University of Science and Technology of China, Hefei 230026, China}
\affiliation{Hefei National Laboratory, University of Science and Technology of China, Hefei 230088, China.}

\author{Xi-Feng Ren}
\email{renxf@ustc.edu.cn}
\affiliation{CAS Key Laboratory of Quantum Information, University of Science and Technology of China, Hefei 230026, China}
\affiliation{Center For Excellence in Quantum Information and Quantum Physics, University of Science and Technology of China, Hefei 230026, China}
\affiliation{Hefei National Laboratory, University of Science and Technology of China, Hefei 230088, China.}

\author{Guang-Can Guo}
\affiliation{CAS Key Laboratory of Quantum Information, University of Science and Technology of China, Hefei 230026, China}
\affiliation{Center For Excellence in Quantum Information and Quantum Physics, University of Science and Technology of China, Hefei 230026, China}
\affiliation{Hefei National Laboratory, University of Science and Technology of China, Hefei 230088, China.}

\date{\today}

\begin{abstract}
Photon-number resolving detectors with hundreds of pixels are now readily available, while the characterization of these detectors using detector tomography is computationally intensive. Here, we present a modified detector tomography model that reduces the number of variables that need optimization. To evaluate the effectiveness and accuracy of our model, we reconstruct the photon number distribution of optical coherent and thermal states using the expectation-maximization-entropy algorithm. Our results indicate that the fidelity of the reconstructed states remains above 99\%, and the second and third-order correlations agree well with the theoretical values for a mean number of photons up to 100. We also investigate the computational resources required for detector tomography and find out that our approach reduces the solving time by around a half compared to the standard detector tomography approach, and the required memory resources are the main obstacle for detector tomography of a large number of pixels. Our results suggest that detector tomography is viable on a supercomputer with 1~TB RAM for detectors with up to 340 pixels.
\end{abstract}

\maketitle


\section{Introduction}
Photon-number resolving (PNR) is important in many classical optics applications, such as X-ray astronomy~\cite{Holland1999} and lidar~\cite{Huang2014}, as well as in quantum optics applications, including quantum random-number generation~\cite{Ren2011}, multiphoton interference~\cite{Feng2023}, high-order correlation measurement~\cite{Dynes2011}, linear optics quantum computation~\cite{Knill2001}, Gaussian boson sampling~\cite{Kaneda2019,Zhong2020}, generation of non-Gaussian quantum states~\cite{Namekata2010}, quantum communication~\cite{Horikiri2006, Sangouard2007} and quantum metrology~\cite{VonHelversen2019,Wang2019,Couteau2023}. There are two main approaches to realize PNR detectors. One is based on the intrinsic PNR capability, such as transition edge sensors~\cite{Fukuda2011}, superconducting nanowire single photon detectors (SNSPDs)~\cite{Cahall2017,Zhu2020}, and avalanche photodiodes~\cite{Kardynal2008}. However, these detectors can only resolve a few photons. The other one is based on multiplexing, which can be further classified according to the dimension into two categories, the one based on bulk optics~\cite{Achilles2003, Fitch2003, Hlousek2019} and the one based on integration~\cite{Divochiy2008, Zhou2014, Mattioli2016}. The multiplexing scheme based on bulk optics suffers from low efficiency, stability and scaling problems due to their large dimensions. It is critical to realize an on-chip integrated PNR detector, which allows the cointegration of spatial or spectral optical components, to scale up the pixels~\cite{Pelucchi2021}.

Among these approaches, SNSPDs hold great potential for on-chip integrated PNR detectors due to their excellent properties, such as high detection efficiency, low dark count rate, high repetition rate and low timing jitter~\cite{You2020,Korzh2020,Lita2022}. Large-scale SNSPD arrays have already been demonstrated~\cite{Wollman2019,Yabuno2020}. Waveguide integrated PNR detectors based on SNSPDs, which are crucial for quantum photonics applications, were reported a decade ago and are capable of resolving up to 4 photons~\cite{Sahin2013}. Recently, a waveguide integrated space-multiplexed PNR detector based on a series of 100 superconducting nanowires~\cite{Cheng2023} as well as a PNR detector based on multiplexing 3 TES detectors~\cite{Eaton2023} have been reported, which is a great breakthrough in the field of PNR detectors and may boost a variety of quantum optics applications in the mesoscopic regime.

Due to the nonideality of the practical optical circuits for multiplexing and the imperfect single-pixel detector elements, one needs to perform quantum detector tomography~\cite{Lundeen2009,Feito2009,Natarajan2013,Endo2021} to characterize PNR detectors. However, the computational resources are demanding when the degrees of freedom are large~\cite{Schapeler2021}. In this work, we propose a modified detector tomography approach that reduces the number of variables to be optimized, while ensuring comparable accuracy to the conventional method. To verify the effectiveness and accuracy of our approach, we numerically reconstruct the photon number distributions of the incident signals using the expectation-maximization-entropy algorithm~\cite{Hlousek2019}, and investigate the needed computational resources. The solving time of our approach is reduced by about a half compared to that of standard detector tomography. The finite memory resource is shown to be the main obstacle for both modified and standard detector tomography approaches. Our results suggest that detector tomography is still feasible for a detector with up to 340 pixels on a supercomputer with 1\,TB RAM.

\section{System configuration}
\begin{figure}[ht]
\centering\includegraphics[width=.5\textwidth]{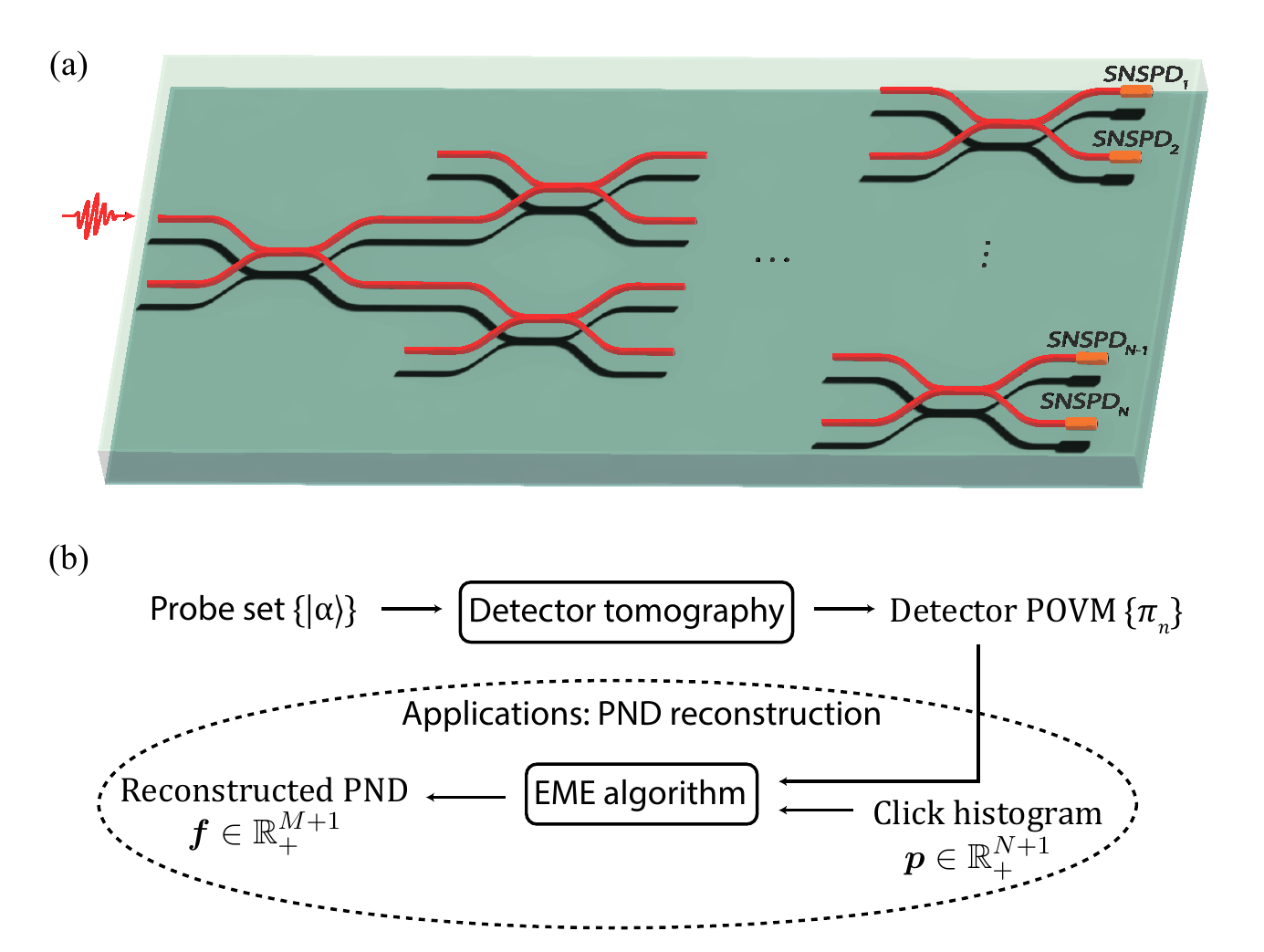}
\caption{The on-chip PNR detector and photon statistics extraction process. (a) Schematic of an on-chip PNR detector consisting of waveguide beam splitters and $N$ SNSPDs. (b) The workflow to reconstruct the PND of an input pulse from the click statistics. A probe set $\{\ket{\alpha}\}$ of coherent states is first used for detector tomography, and the POVM elements $\{\pi_n\}$ are obtained. The PND $\vb*{f}\in\mathbb{R}_+^{M+1}$ of the input state, where $\mathbb{R}_+$ represents nonnegative real numbers and $M$ is the photon number at which the Fock space is truncated, can be reconstructed by the EME algorithm from the click statistics $\vb*{p}\in\mathbb{R}_+^{N+1}$ measured by the PNR detector. PNR: photon-number resolving, SNSPD: superconducting nanowire single-photon detector, PND: photon-number distribution, POVM: positive operator-valued measurement, EME: expectation-maximization-entropy.}\label{fig1}
\end{figure}

\begin{figure}[ht]
\centering\includegraphics[width=.35\textwidth]{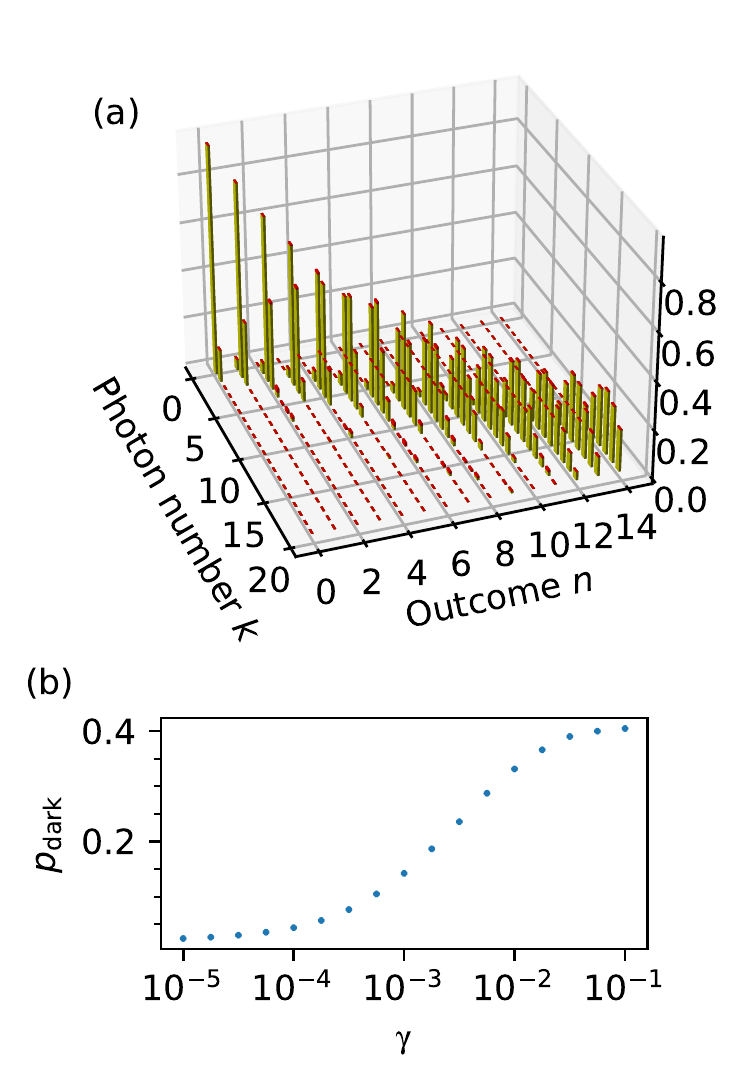}
\caption{An example of detector tomography. (a) The reconstructed POVM elements of a 70-pixel detector with regularization parameter $\gamma=10^{-4}$. The POVM elements obtained from modified detector tomography are shown as yellow bars. The red bars represent the absolute differences between POVM elements from modified and standard detector tomography approaches. Only the first 15 POVM elements are shown for clarity. (b) The dark-count probability extracted from the reconstructed POVM elements with respect to the regularization parameter $\gamma$.}\label{fig2}
\end{figure}

A photonic-integrated circuit-based on-chip PNR detector employing SNSPDs is schematically illustrated in Fig. \ref{fig1}(a). By a spatial-multiplexing photonic circuit, the photons in an input pulse are nearly evenly distributed to $N$ SNSPDs with similar detection efficiency. An alternate approach (not shown) is to integrate an array of SNSPDs on a single waveguide~\cite{Cheng2023,Sahin2013} such that the absorption efficiency of each single pixel is designed to absorb the input photons with nearly equal probability. Apart from the capability of resolving photon numbers~\cite{Li2020}, SNSPDs are compatible with other photonic components on photonic chips~\cite{Pernice2012}, thus allowing a variety of applications, ranging from single-photon spectrometers~\cite{Cheng2019}, on-chip Boson sampling~\cite{Spring2013}, to hybrid quantum chips for quantum information processing~\cite{Xu2022}.

To evaluate the performances of detector tomography with hundreds of pixels, we implement numerical Monte Carlo simulations in the following, where the code can be found in Ref~\cite{Liu2023}. We set the device configuration with the coupling efficiency from the input to the detector to be $99\%$ to account for potential device insertion loss. Considering the fabrication imperfections, the input photons are nearly evenly distributed to each pixel with a relative uncertainty of 2\%. The simulations assume no dark counts and each detector pixel has a uniformly distributed intrinsic detection efficiency (the probability of generating clicks when a photon is absorbed by the pixel) between $90\%$ and $95\%$~\cite{Zhang2017}. Note that the path-dependent propagation losses could also be included in the absorption efficiency or intrinsic efficiency of individual pixels. The technical noise of the laser~\cite{Lundeen2009}, which is used to generate the probe set $\{\ket{\alpha}\}$, is also considered by assuming that the mean number of photons of each pulse is normally 
distributed as $\abs{\beta}^2\sim\mathcal{N}(\mu=\abs{\alpha}^2, \sigma=0.0188\abs{\alpha}^2)$.

\section{Detector tomography}
Due to the fabrication imperfections of photonic circuit components and superconducting devices, PNR detectors are usually unbalanced and there may be cross-talk between pixels. Quantum detector tomography~\cite{Lundeen2009,Feito2009,Natarajan2013,Endo2021} aims to characterize PNR detectors by determining the positive operator-valued measurement (POVM) elements. The PNR detectors without phase dependence can be described by the POVM diagonal in the Fock state basis as
\begin{equation}
    \pi_n = \sum_{k=0}^M\theta^{(n)}_k\dyad{k},\quad n=0,1,\dots,N,
\end{equation}
with element $\pi_n$ corresponding to the outcome of $n$ clicks of an $N$-pixel detector, and the Fock space is truncated at a photon number of $M$. As shematically illustrated in Fig.~\ref{fig1}(b), the obtained detector POVM can be applied to reconstruct the photon number distribution (PND) of the input state by only providing the measured statistic of detector clicks. In particular, the PND vector $\vb*{f}\in\mathbb{R}_+^{M+1}$ of the incident signal, where $\mathbb{R}_+$ represents nonnegative real numbers, can be reconstructed from the measured statistics $\vb*{p}\in\mathbb{R}_+^{N+1}$ using the  expectation-maximization-entropy (EME) algorithm~\cite{Hlousek2019}.

\subsection{Standard detector tomography}
A set of $D$ coherent states $\{\ket{\alpha}\}$ with different mean numbers $\abs{\alpha}^2$ of photons are used as probe states for detector tomography, and the corresponding click statistics are obtained to reconstruct the POVM elements by solving the following convex optimization problem~\cite{Lundeen2009,Feito2009}:
\begin{equation}\label{detector tomography}
    \begin{aligned}
        \mathrm{min} \quad &{\Vert P-F\varPi \Vert}_{\mathrm{Fro}} + \tilde{\gamma}\sum_{n,k}[\theta^{(n)}_k - \theta^{(n)}_{k+1}]^2, \\
        \mathrm{s.t.} \quad &\pi_n \ge 0, \quad \sum_n \pi_n = I.
    \end{aligned}
\end{equation}
Here, $\|A\|_{\mathrm{Fro}}=(\sum_{i,j}A_{i,j}^2)^{1/2}$ represents the Frobenius norm, $P\in\mathbb{R}_+^{D\times(N+1)}$ is a matrix containing the measured statistics of the probe states, and $F\in\mathbb{R}_+^{D\times(M+1)}$ is a matrix containing the probe states $\{\ket{\alpha}\}$. Each row of $P$ and $F$ corresponds to the measured statistics $\vb*{p}$ and the true PND $\vb*{f}$ of a probe state, respectively. $\varPi\in\mathbb{R}_+^{(M+1)\times(N+1)}$ is a matrix containing the $N+1$ POVM elements where $\varPi_{kn}=\theta^{(n)}_k$. A regularization parameter $\tilde{\gamma}$ is introduced to suppress ill-conditioning and noise.

To fully characterize the response of the detector with respect to input states, the maximum mean number of photons of the probe states, denoted as $\abs{\alpha}^2_{\mathrm{max}}$, should be chosen such that the probability that all the $N$ pixels click simultaneously saturates~\cite{Feito2009}. In our Monte Carlo simulation, we choose $\abs{\alpha}^2_{\mathrm{max}}$ such that the probability of measuring more than $N$ clicks by PNR detector is greater than $90\%$ when the mean input photon number $\ev{n}=\abs{\alpha}^2_{\mathrm{max}}$. The sample set of the input probe states is selected as the coherent states with the mean photon number $\ev{n}$ ranging from 1 to $|\alpha|^2_\mathrm{max}$ in steps of 1. We choose the truncation parameter $M>\abs{\alpha}^2_{\mathrm{max}}$ such that the probability of the Poisson distribution at $M$ satisfies the condition $(\abs{\alpha}^2_{\mathrm{max}})^Me^{-\abs{\alpha}^2_{\mathrm{max}}}/M!\le 10^{-5}$. 
For each probe state, $10^5$ sample pulses are used to obtain the measured statistics $\vb*{p}$.

\subsection{Modified detector tomography}

\begin{figure}
    \centering\includegraphics[width=.45\textwidth]{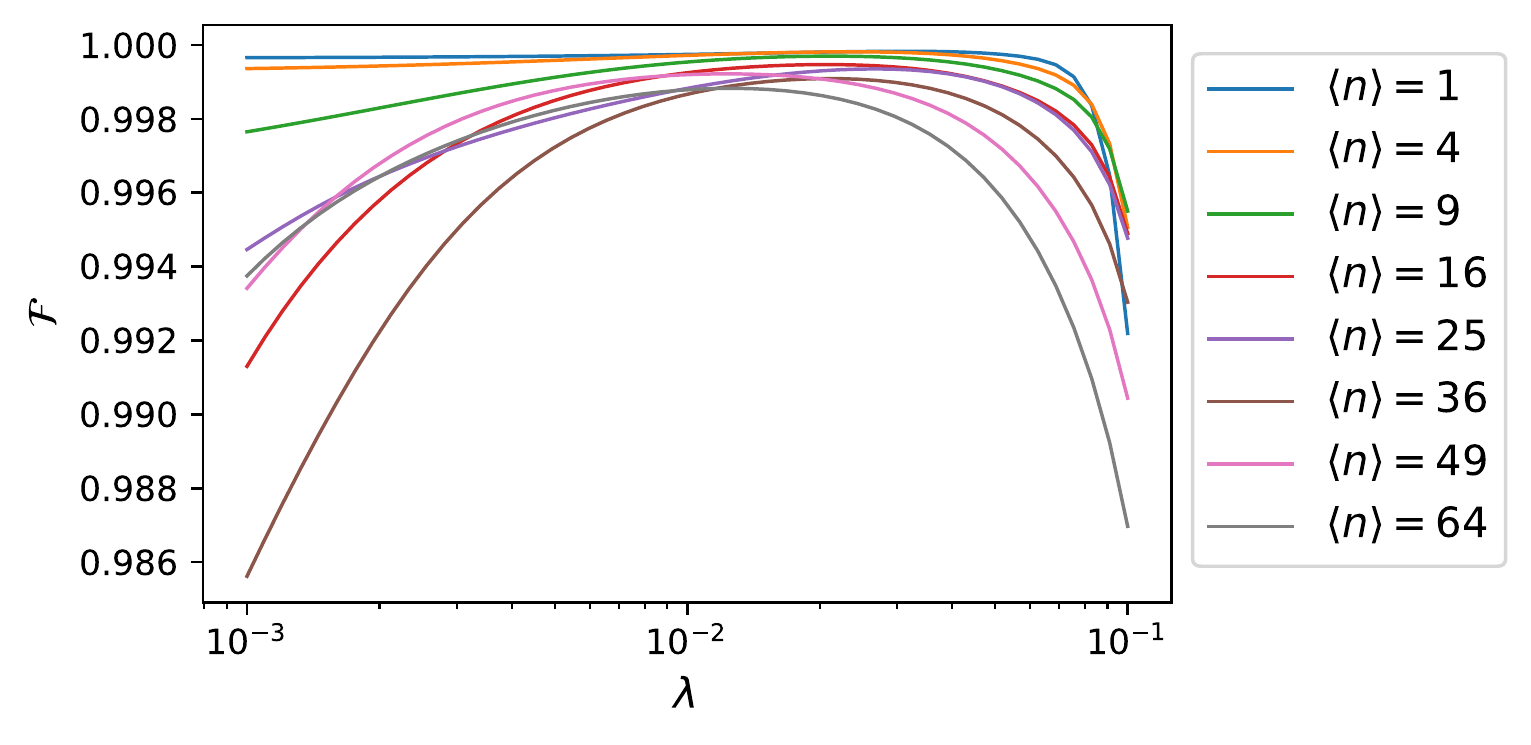}
    \caption{The fidelity $\mathcal{F}$ of the reconstructed PNDs with respect to the regularization parameter $\lambda$ for input coherent states with various mean photon numbers $\ev{n}$.
    }\label{fig3}
\end{figure}

\begin{figure*}
    \centering\includegraphics[width=0.8\textwidth]{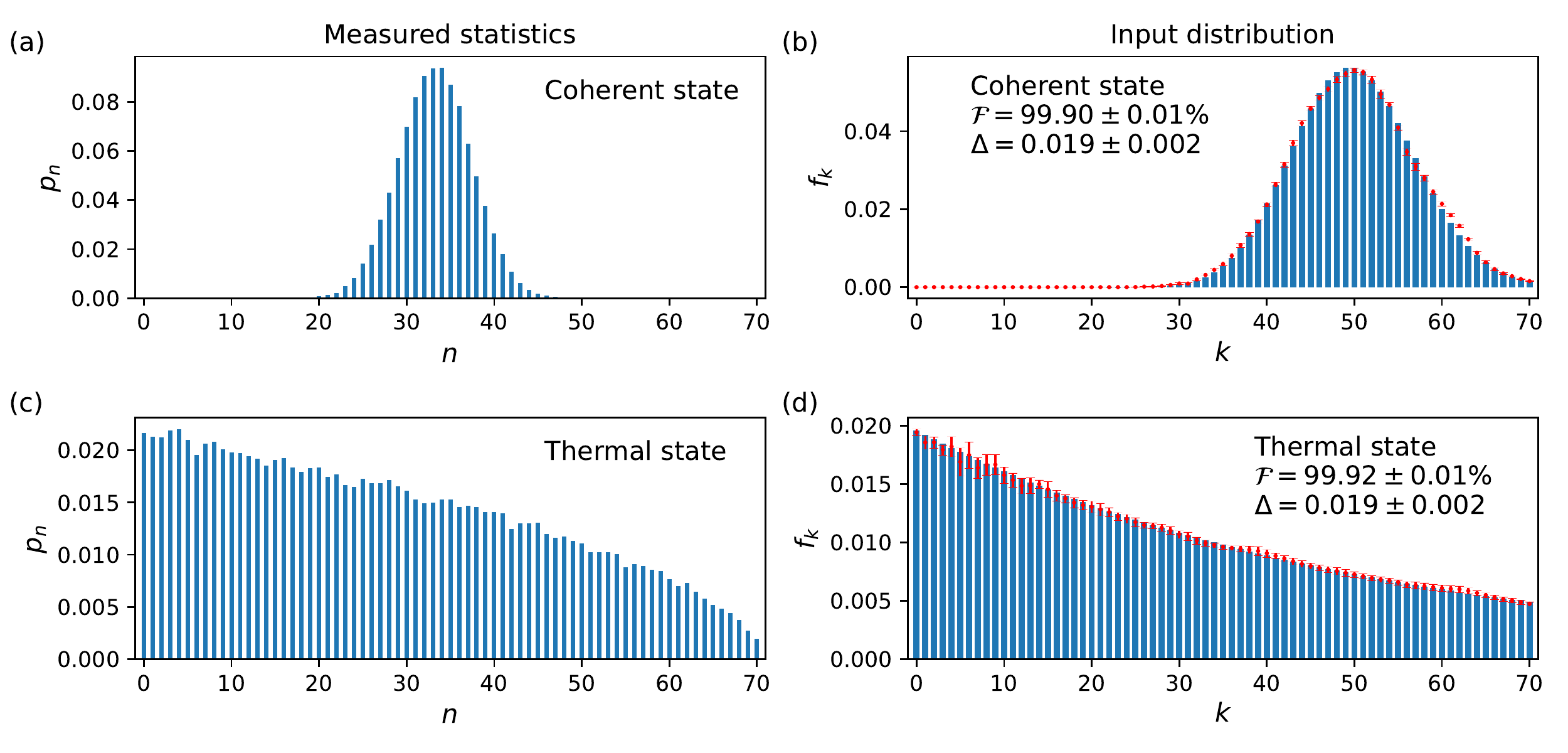}
    \caption{(a) and (b) The measured statistics of clicks and reconstructed photon number distributions for coherent state input, with an input mean photon number  $\ev{n}=50$. 
    (c) and (d) The measured statistics of clicks and reconstructed photon number distributions for thermal state input, with an input mean photon number  $\ev{n}=50$. 
    The error bars are the standard deviations calculated by repeating the click statistics and state reconstruction process 10 times.
    }\label{fig4}
\end{figure*}

For an $N$-pixel detector with input state space truncated at photon number $M,(M>N)$, the number of variables to be optimized is $(M+1)\times(N+1)$ , which is on the order of $10^4$ for a 100-pixel detector. Due to the large degrees of freedom of the POVM elements, the resources needed to perform detector tomography are demanding~\cite{Schapeler2021}. In this section, we derive a modified form of detector tomography, which would reduce the number of variables to be optimized. 

To make the objective function in Eq. \eqref{detector tomography} differentiable, we change the form of the convex problem to
\begin{equation}\label{norm squared}
    \begin{aligned}
        \mathrm{min}\quad &\frac{1}{2}\|P-F\varPi\|_{\mathrm{Fro}}^2 + \frac{\gamma}{2}\sum_{k=0}^{M-1}\sum_{n=0}^{N}(\varPi_{k,n}-\varPi_{k+1,n})^2 \\
        \mathrm{s.t.}\quad &\varPi\boldsymbol{1}_{N+1}=\boldsymbol{1}_{M+1}, \\
        &\varPi_{k,n}\ge 0, \quad k=0,\dots,M;\ n=0,\dots,N;
    \end{aligned}
\end{equation}
where $\gamma$ is the regularization parameter and $\boldsymbol{1}_{N+1}$ is a $(N+1)$-dimensional vector with all components being one.

Denote $\vb*{u}_k\equiv\vb*{e}_k-\vb*{e}_{k+1}$, where $\vb*{e}_k\in\mathbb{R}^{M+1}$ is the $k$th basis vector. Then the objective function can be written as
\begin{equation}
    \begin{aligned}
        f(\varPi) &\equiv \frac{1}{2}\|P-F\varPi\|_{\mathrm{Fro}}^2 + \frac{\gamma}{2}\sum_{k=0}^{M-1}\sum_{n=0}^{N}(\varPi_{k,n}-\varPi_{k+1,n})^2 \\
        &= \frac{1}{2}\mathrm{tr}[(P-F\varPi)^\mathrm{T}(P-F\varPi)] + \frac{\gamma}{2}\sum_{k=0}^{M-1}\vb*{u}_k^\mathrm{T}\varPi\varPi^\mathrm{T}\vb*{u}_k \\
        &= \frac{1}{2}\mathrm{tr}[(P-F\varPi)^\mathrm{T}(P-F\varPi)] + \frac{\gamma}{2}\mathrm{tr}[U\varPi\varPi^\mathrm{T}],
    \end{aligned}
\end{equation}
  where
\begin{equation}
    U = \sum_{k=0}^{M-1}\vb*{u}_k\vb*{u}_k^\mathrm{T}.
\end{equation}

The gradient of the objective function is
\begin{equation}
    \begin{aligned}
        \nabla f(\varPi) &= -F^\mathrm{T}(P-F\varPi) + \gamma U\varPi \\
        &= -F^\mathrm{T}P + \left(F^\mathrm{T}F + \gamma U\right)\varPi,
    \end{aligned}
\end{equation}
and the corresponding solution to $\nabla f(\varPi)=0$ is
\begin{equation}
    \widetilde{\varPi} = (F^\mathrm{T}F + \gamma U)^{-1}F^\mathrm{T}P.
\end{equation}
Since $\nabla f(\widetilde{\varPi})=0$, we have $\nabla f(\widetilde{\varPi})\boldsymbol{1}_{N+1}=0$, i.e.,
\begin{equation}
    \left[-F^\mathrm{T}P + \left(F^\mathrm{T}F + \gamma U\right)\widetilde{\varPi}\right]\boldsymbol{1}_{N+1} = 0.
\end{equation}
Based on the fact that $P\boldsymbol{1}_{N+1}=\boldsymbol{1}_D$, the last equation becomes
\begin{equation}
    \left(F^\mathrm{T}F + \gamma U\right)\widetilde{\varPi}\boldsymbol{1}_{N+1} = F^\mathrm{T}\boldsymbol{1}_{D},
\end{equation}
which holds when
\begin{equation}
    \begin{aligned}
        F\widetilde{\varPi}\boldsymbol{1}_{N+1} &= \boldsymbol{1}_{D}, \\
        \gamma U\widetilde{\varPi}\boldsymbol{1}_{N+1} &= 0.
    \end{aligned}
\end{equation}
Therefore, $\nabla f(\widetilde{\varPi})=0$ leads to
\begin{equation}
    \widetilde{\varPi}\boldsymbol{1}_{N+1} = \boldsymbol{1}_{M+1},
\end{equation}
which indicates that $\widetilde{\varPi}$ satisfies the equality constraints in Eq. \eqref{norm squared}. However, nearly half of the inequality constraints in Eq. \eqref{norm squared} does not hold for $\widetilde{\varPi}$ in our simulation. To simplify the solving process of detector tomography, intuitively we introduce a treatment that  sets $\varPi_{k,n}=0$ if $\widetilde{\varPi}_{k,n}\le 0$. This approximation reduces the number of variables by about half, which leads to a decrease in solving time, as will be demonstrated in Section \ref{sec:performance}. Following this treatment, detector tomography can be reformulated as
\begin{equation}\label{MDT}
    \begin{aligned}
      \mathrm{min}\quad &\frac{1}{2}\|P-F\varPi\|_{\mathrm{Fro}}^2 + \frac{\gamma}{2}\sum_{k=0}^{M-1}\sum_{n=0}^{N}(\varPi_{k,n}-\varPi_{k+1,n})^2 \\
      \mathrm{s.t.}\quad &\varPi\boldsymbol{1}_{N+1}=\boldsymbol{1}_{M+1}, \\
      &\varPi_{k,n}=0, \ \mathrm{if}\ \widetilde{\varPi}_{k,n}\le0, \\
      &\varPi_{k,n}\ge 0, \ \mathrm{if}\ \widetilde{\varPi}_{k,n}>0.
    \end{aligned}
\end{equation}
We refer to Eq. \eqref{MDT} as modified detector tomography (MDT) and Eq. \eqref{norm squared} or \eqref{detector tomography} as standard detector tomography (SDT).

We solve Eq. \eqref{MDT} and Eq. \eqref{norm squared} using CVXPY~\cite{Diamond2016, Agrawal2018} with a commercial solver called MOSEK which supports multi-threading~\cite{mosek}. The solution to MDT coincides with that of SDT within acceptable accuracy (relative error less than $3\%$). As an example, the reconstructed POVM elements of a 70-pixel PNR detector by detector tomography are shown in Fig.~\ref{fig2}(a), where the Fock space is truncated at $M=608$ and the regularization parameter is chosen as $\gamma=10^{-4}$. The yellow bars show the results for MDT, while the red bars indicate the absolute difference of POVM elements obtained by MDT and SDT. The agreement between the MDT and SDT confirms that the MDT approximates SDT quite well.

Note that the regularization parameter should be chosen below a certain threshold such that the dark-count probability extracted from the reconstructed POVM elements coincides with that measured experimentally~\cite{Schapeler2020}, as shown in Fig. \ref{fig2}(b). The dark-count probability defined as the single-click probability when no photons are incident is $p_{\mathrm{dark}}=\varPi_{0,1}=4.4\%$ for $\gamma=10^{-4}$. For a larger $\gamma$, the dark-count probability deviates from the assumption of no dark counts in the simulation, and for smaller $\gamma$, errant spikes occur in the reconstructed POVM elements in their distribution in photon numbers.

\section{Verification: reconstruction of photon number distribution}

To verify the effectiveness and accuracy of the POVM elements obtained from the MDT model, we numerically tested its performance for a 70-pixel PNR detector, with the procedure following the schematic in Fig.~\ref{fig1}(b). We reconstruct the PNDs of both coherent and thermal states using the EME algorithm~\cite{Hlousek2019}. The algorithm works by iterating the following equations:
\begin{subequations}
    \begin{gather}
        f_k^{(i+1)} = R^{(i)}_kf_k^{(i)} -\lambda(\ln f_k^{(i)}+S^{(i)})f_k^{(i)}, \\
        R^{(i)}_k = \sum_{n=0}^N \frac{p_n}{\sum_{k'=0}^M\varPi_{k'n}f^{(i)}_{k'}}\varPi_{kn},\\
        \quad S^{(i)} = -\sum_{k=0}^{M} f^{(i)}_k\ln f^{(i)}_k,
    \end{gather}
\end{subequations}
where the superscript $(i)$ represents the $i$th iteration and $\lambda$ is the regularization parameter. The initial guess $\vb*{f}^{(0)}$ of the input distribution is set to be uniform.

\begin{figure}
\centering\includegraphics[width=.45\textwidth]{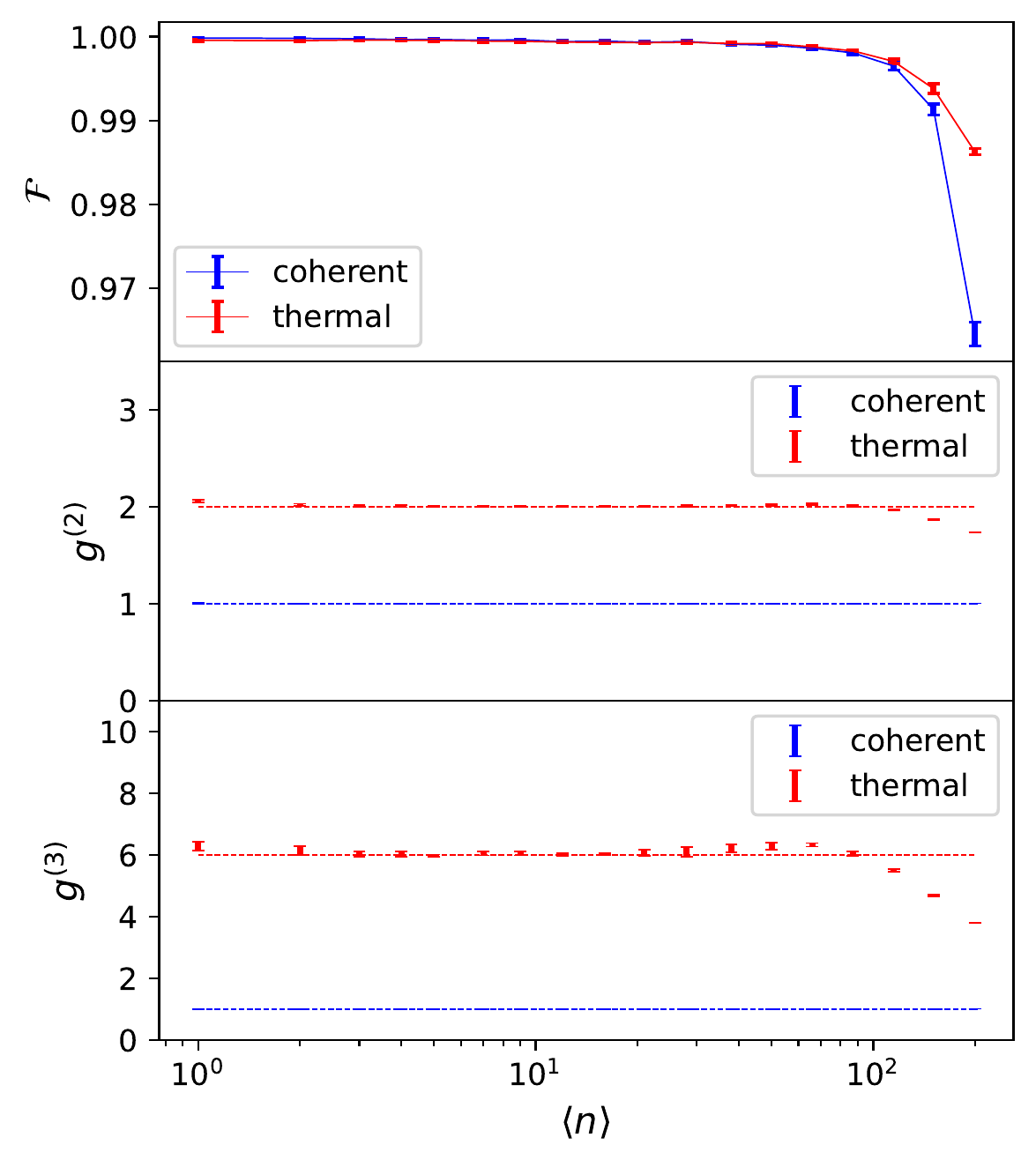}
\caption{The fidelity $\mathcal{F}$ and correlation functions $g^{(2)}$ and $g^{(3)}$ of the reconstructed PNDs of coherent states and thermal states with mean photon number $\ev{n}$. The dashed lines for the correlation functions are the corresponding theoretical values, i.e., $g^{(2)}=1(2)$ and $g^{(3)}=1(6)$ for coherent (thermal) states.}\label{fig5}
\end{figure}

First, we choose a regularization parameter $\lambda$ for the EME algorithm. By using the probe states in the previous section, the fidelity of reconstruction $\mathcal{F}=(\sum_{k=0}^M\sqrt{f_kf_k^{\mathrm{true}}})^2$ is evaluated for different $\lambda$. As shown in Fig.~\ref{fig3}, the fidelities achieve the highest value when $\lambda=0.02$, so we set $\lambda=0.02$ in the following reconstruction processes. We should note that since the maximum entropy regularization has an effect of smoothing the distribution, it cannot be used to reconstruct states that are not ``smooth'', such as squeezed vacuum states and Fock states. For such states, prior information is necessary for accurate reconstruction of PNDs.

Figure~\ref{fig4} shows typical results of the PND reconstruction for the coherent and thermal input states with $\ev{n}=50$. In Fig.~\ref{fig4}(a) and (c), the measured statistics of clicks for the coherent and thermal states are plotted. The measured distributions have slight fluctuations that are different from the smooth distributions, which are attributed to the finite sample numbers. We also notice that the measured statistics of the thermal state in Fig. \ref{fig4}(c) deviate from a geometric distribution as predicted for an ideal thermal state, which indicates that the statistic of clicks could be significantly changed by PNR detectors with a finite number of pixels of the detector and a reliable reconstruction of PND is necessary.

The corresponding theoretical PNDs for the input states are shown by blue bars in Figs~\ref{fig4}(b) and (d), where only the first 70 components are shown. With our method, we reconstructed the PNDs from the measured statistics and the optimized POVM of the PNR detector, and the results are shown by red circles in Fig.~\ref{fig4}(b) and (d).
The results are obtained by numerically repeating the click statistics simulation and state reconstruction process 10 times, and the corresponding standard deviations of the reconstructed PNDs are also shown as error bars.
The fidelity is above $99.9\%$ and the total variation distance $\Delta=\sum_{k=0}^M\abs{f_k-f^{\mathrm{true}}_k}/2$ is also calculated and shown in the figure, which indicates the high accuracy of our reconstruction method. 

Figure~\ref{fig5} further evaluated the performance of our approach by calculating the fidelity $\mathcal{F}$ and the high-order photon correlation functions $g^{(2)}$ and $g^{(3)}$ for the reconstructed coherent and thermal states for different $\ev{n}$. We find that our approach can reconstruct both coherent and thermal states with high fidelities $\mathcal{F}>99\%$ even when $\ev{n}$ approaches 100. For the $g^{(2)}$ function, our results agree with the theoretical predictions of $2$ and $1$ for thermal and coherent states, respectively. Similarly, the reconstructed $g^{(3)}\approx6$ and $1$ agree with theory. The fidelity drops and the correlation functions deviate from the theoretical values as $\ev{n}$ increases above 100, because the probability of more than one photon entering the same pixel becomes nonnegligible when $\ev{n}>N$. These results validate the MDT model and imply the potential of our approach for reconstructing PNDs of PNR detectors accurately even when the mean photon number exceeds 100.

\section{\label{sec:performance}Performance evaluation}
\begin{figure}
\centering\includegraphics[width=.5\textwidth]{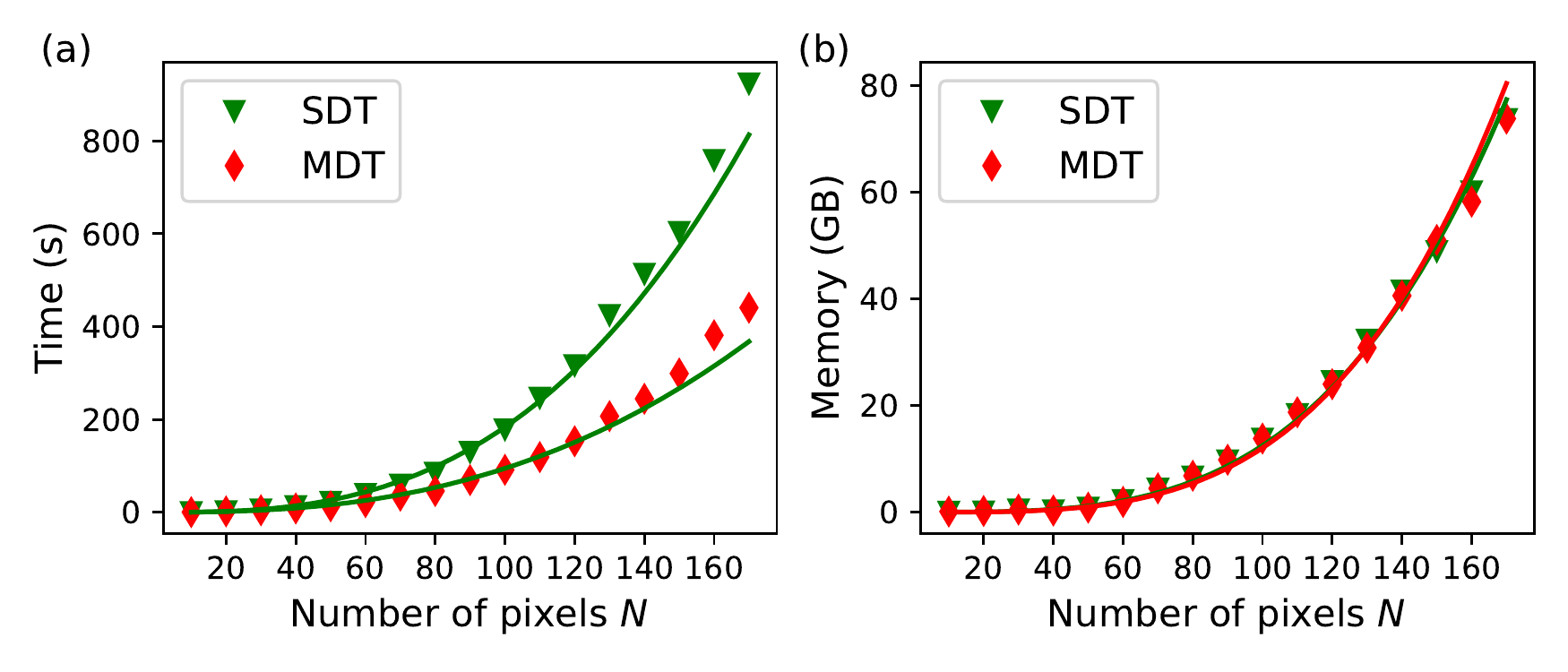}
\caption{The computational cost of detector tomography. The solving time (in seconds) and memory consumption (in GB) with respect to the different number of pixels for SDT (triangle) and MDT (diamonds) were profiled using solver MOSEK with 16 threads. The solid curves are the corresponding fittings.}\label{fig6}
\end{figure}

Although it is known that detector tomography is computationally demanding when the number of pixels or the truncated dimension of the Fock state space increases, the limit of the detector pixels has not been explored. Here, we numerically characterize the time and memory consumption for detector tomography with respect to the number of pixels on a server with two 3.2\,GHz Xeon E5-2667 CPUs and 128\,GB RAM, as shown in Fig.~\ref{fig6}. Convex optimization problem are known to be solved efficiently in polynomial time~\cite{Boyd2004}, and the time and memory consumption with respect to the number of pixels are fitted with the weighted least square method by using the model $y=aN^b$. We obtain
\begin{subequations}
    \begin{align}
        t_{\mathrm{SDT}} &= 4.51\times 10^{-4}\times N^{2.80} \ \mathrm{s}, \\
        m_{\mathrm{SDT}} &= 1.66\times 10^{-6}\times N^{3.44} \ \mathrm{GB} 
    \end{align}
\end{subequations}
 gives for SDT, and
\begin{subequations}
    \begin{align}
        t_{\mathrm{MDT}} &= 7.17\times 10^{-4}\times N^{2.56} \ \mathrm{s},\\
        m_{\mathrm{MDT}} &= 0.85\times 10^{-6}\times N^{3.58} \ \mathrm{GB}
    \end{align}
\end{subequations}
for MDT. Note that we only consider the time spent by the solver and that the time spent by CVXPY for compiling the problem is not taken into consideration. In our numerical simulations, the variables (degrees of freedom for optimization) of MDT are reduced by about 40\%, which leads to a decrease in solving time by approximately half compared to that of SDT, as shown in Fig.~\ref{fig6}(a), while the memory consumptions for MDT and SDT [Fig.~\ref{fig6}(b)] are comparable. The reason may be attributed to the number of constraints being the same for both models, and it requires similar memory to compile these two models in CVXPY. 

These results indicate that the main obstacle of detector tomography in practice is finite memory resources. Our results suggest that for detectors with similar efficiency and dark-count probability as in our simulation, by employing the currently feasible supercomputer with 1 TB RAM, the modified detector tomography can handle PNR detectors with up to 340 pixels.

\section{Conclusion}
In conclusion, we propose a modified detector tomography approach that reduces the degrees of freedom without sacrificing precision. The solution obtained using this method coincides with that of standard detector tomography, with a relative error of less than 3\%.
As a verification of the effectiveness and accuracy of the MDT model, we reconstruct photon number distributions of coherent and thermal states using expectation-maximization-entropy algorithm for a 70-pixel photon number resolving detector. The fidelity of the reconstructed states remains above 99\% and the second and third order coherence $g^{(2)}, g^{(3)}$ agrees well with the theoretical values for $\ev{n}$ up to 100. In addition, we also provide insights into the computational constraints associated with multipixel detector tomography. The solving time of our modified detector tomography is shown to be nearly 2 times shorter than that of standard detector tomography, and the main obstacle for detector tomography is the finite memory resource.
For detectors with comparable efficiency and dark-count probability to that in our simulation, we suggest that the number of pixels of around 340 is manageable with available computer resources (supercomputer with 1 TB RAM).

\begin{acknowledgments}
This work was funded by the National Natural Science Foundation of China (Grants Nos.~62061160487, 92265210, 12061131011, 11922411), the Major Scientific Project of Zhejiang Laboratory (No.~2020LC0AD01), and the Innovation Program for Quantum Science and Technology (No.~2021ZD0303200), and the Key Research and Development Program of Anhui Province (2022b1302007). CLZ was also supported by the Fundamental Research Funds for the Central Universities, USTC Research Funds of the Double First-Class Initiative. The work is also supported by the supercomputing system in the Supercomputing Center of USTC the USTC Center for Micro and Nanoscale Research and Fabrication.
\end{acknowledgments}

\end{document}